\documentclass[sigconf]{acmart}

\AtBeginDocument{%
  }

\setcopyright{acmlicensed}
\copyrightyear{2025}
\acmYear{2025}
\acmDOI{XXXXXXX.XXXXXXX}

\acmISBN{978-1-4503-XXXX-X/YY/MM}

\usepackage{multirow}
\usepackage{graphicx}
\usepackage{subcaption}

\definecolor{cream}{rgb}{1.0, 0.98, 0.92} 

\begin{document}

\title[GenUP: Generative User Profilers as In-Context Learners for Next POI Recommender Systems]{GenUP: Generative User Profilers as In-Context Learners for Next POI Recommender Systems}

\author{Wilson Wongso}
\orcid{0000-0003-0896-1941}
\affiliation{%
  \institution{University of New South Wales}
  \city{Sydney}
  \country{Australia}
}
\email{w.wongso@unsw.edu.au}

\author{Hao Xue}
\orcid{0000-0003-1700-9215}
\affiliation{%
  \institution{University of New South Wales}
  \city{Sydney}
  \country{Australia}
}
\email{hao.xue1@unsw.edu.au}

\author{Flora D. Salim}
\orcid{0000-0002-1237-1664}
\affiliation{%
  \institution{University of New South Wales}
  \city{Sydney}
  \country{Australia}
}
\email{flora.salim@unsw.edu.au}

\renewcommand{\shortauthors}{Wongso et al.}

\begin{abstract}
Traditional Point-of-Interest (POI) recommendation systems often lack transparency, interpretability, and scrutability due to their reliance on dense vector-based user embeddings. Furthermore, the cold-start problem---where systems have insufficient data for new users---limits their ability to generate accurate recommendations. Existing methods often address this by leveraging similar trajectories from other users, but this approach can be computationally expensive and increases the context length for LLM-based methods, making them difficult to scale. To address these limitations, we propose a method that generates natural language (NL) user profiles from large-scale, location-based social network (LBSN) check-ins, utilizing robust personality assessments and behavioral theories. These NL profiles capture user preferences, routines, and behaviors, improving POI prediction accuracy while offering enhanced transparency. By incorporating NL profiles as system prompts to LLMs, our approach reduces reliance on extensive historical data, while remaining flexible, easily updated, and computationally efficient. Our method is not only competitive with other LLM-based methods but is also more scalable for real-world POI recommender systems. Results demonstrate that our approach consistently outperforms baseline methods, offering a more interpretable and resource-efficient solution for POI recommendation systems. Our source code is available at: \url{https://github.com/w11wo/GenUP/}.
\end{abstract}



\keywords{Large language models, Location-based social networks, Point-of-interest recommendation}


\maketitle

\section{Introduction}

Location-based social networks (LBSNs) generate vast amounts of data, offering significant potential for modeling human mobility patterns. One application of LBSN data is next Point-of-Interest (POI) recommendation, where the goal is to predict a user’s likely next destination based on their historical check-ins. POI recommendation systems have evolved from early Markov models \cite{cheng2013you,rendle2010factorizing} to more advanced deep learning techniques, such as sequential models \cite{kong2018hst,sun2020go} and Graph Neural Networks \cite{lim2020stp,zhang2022next,rao2022graph,yang2022getnext}. These models have significantly improved the ability to capture spatial-temporal dependencies and user mobility patterns. However, challenges remain, particularly with the cold-start problem, where limited interaction data makes it difficult to provide accurate predictions for inactive users. This is typically resolved by learning from similar historical trajectories from other users. More recently, large language models (LLMs) have been applied to next POI recommendation \cite{wang2023would,beneduce2024large}, leveraging their commonsense reasoning capabilities and their ability to incorporate contextual information. However, LLM-based methods can be computationally intensive due to their reliance on long historical trajectory prompts to capture cross-user patterns \cite{li2024large}. Similarly, agentic LLM frameworks \cite{feng2024agentmove}, which employ multiple agents for different tasks introduce added complexity and resource demands.

Another challenge in POI recommendation is personalization, often achieved through dense vector representations of users. While these embeddings effectively model user behavior, they lack interpretability, transparency, and scrutability—qualities crucial for building user trust and improving recommendation explainability \cite{10.1145/1040830.1040870,INR-066}. Natural language (NL) user profiles offer a promising alternative, where new users can quickly provide a concise description of their preferences, and existing users can update their profiles for more personalized and transparent recommendations \cite{radlinski2022natural}. Despite their potential, NL profiles have not seen widespread adoption due to difficulties in evaluation and the ongoing development of methods for accurately generating these profiles \cite{ramos2024natural}. Prior work, such as LLMob \cite{wang2024large}, relies on rigid user categorization, generating multiple candidate personas and selecting the most likely one based on a given trajectory. While effective, this approach limits personalization by constraining users to predefined categories. In response, we introduce NL user profiles for next POI recommendation, replacing long historical trajectory prompts with concise profiles to reduce computational overhead and context length. We propose generating these profiles from LBSN check-ins to learn user behaviors and integrate robust behavioral theories for more accurate POI predictions.

\paragraph{Contributions}
We introduce a novel approach for deriving NL user profiles from \textbf{implicit} spatiotemporal signals in check-in data, a domain that has been largely overlooked in prior work. Existing LLM-driven recommender systems predominantly operate on datasets with \textbf{explicit} user feedback (e.g., ratings, reviews), whereas our method infers user preferences and routines from raw LBSN trajectories, requiring higher-level behavioral inference. To the best of our knowledge, we are the first to generate NL user profiles directly from LBSN check-in sequences and integrate them into a next POI recommendation framework. Our approach constructs these profiles using structured prompting that combines user attributes and personality traits, specifically leveraging components from the Theory of Planned Behavior and the Big Five Inventory. Unlike prior prompt-based user modeling methods, which apply generic persona templates or fixed category assignments, our profiles are data-driven, dynamically generated, and domain-specific, tailored to spatiotemporal mobility patterns. Our method enhances LLM-based POI recommendation by improving transparency, reducing reliance on extensive historical data, and maintaining competitive accuracy with a more scalable, resource-efficient design.

\section{Related Works}

\subsection{Large Language Models for Next POI Recommendation}

Significant advancements in mobility prediction have been driven by deep learning models, moving from traditional Markov models \cite{cheng2013you,rendle2010factorizing} to more sophisticated approaches that capture high-order mobility patterns in trajectory data. Early works primarily relied on sequential-based methods like Recurrent Neural Networks (RNNs) and attention mechanisms to model individual user behavior. Models such as HST-LSTM \cite{kong2018hst} and LSTPM \cite{sun2020go} were designed to capture spatio-temporal dependencies but lacked the ability to incorporate cross-user collaborative signals. This limitation prompted the development of graph-based models \cite{lim2020stp,zhang2022next,rao2022graph,yang2022getnext} to model relationships between users, locations, and trajectories. These models addressed the cold-start problem more effectively by using collaborative signals but struggled to integrate commonsense knowledge and broader contextual information \cite{li2024large}.

Large language models (LLMs) have since emerged as a viable solution for next POI recommendation, offering the ability to process both spatial-temporal data and commonsense knowledge. LLM-Mob \cite{wang2023would}, for example, introduced in-context learning and exploited the zero-shot reasoning capabilities of LLMs \cite{wei2022emergent} to enhance next POI recommendation given the historical and contextual trajectories of a user. Similarly, LLM-ZS \cite{beneduce2024large} simplified the zero-shot process and analyzed zero-shot, one-shot, and few-shot prompting scenarios. However, both methods depend heavily on task-specific prompting and require very large LMs, such as 70B instruction-tuned or closed-source LLMs, to achieve competitive performance. AgentMove \cite{feng2024agentmove} extends these techniques by employing multiple LLMs agents \cite{wang2024survey,xi2023rise}, each dedicated to different components of mobility prediction. This framework further integrates external knowledge sources such as urban structures, geospatial data, and location graphs, enabling the predictor LLM to process richer contextual information for the prediction process. While this multi-agent approach improves prediction accuracy and handles cold-start issues, it is computationally intensive and less suitable for real-time or resource-constrained environments.

LLM4POI \cite{li2024large} proposed an alternative by conducting a supervised fine-tuning (SFT) of LLMs for next POI prediction, framing it as a question-answer task. This approach simplified the process by (1) using lighter 7B open-source LLMs, (2) streamlining in-context learning with trajectory prompting tailored for LBSN data, and (3) leveraging trajectory similarity from both current and other users to enhance the LLM’s predictive capabilities. While leveraging cross-user similar trajectories improves accuracy, LLM4POI’s reliance on long context lengths results in high computational costs. Building on this, our method introduces natural language user profiles to personalize POI recommendations and learn cross-user patterns from similar profiles, removing the need for long historical trajectory prompts and significantly decreasing the context length.

\begin{figure*}[ht]
  \centering
  \includegraphics[width=0.85\textwidth]{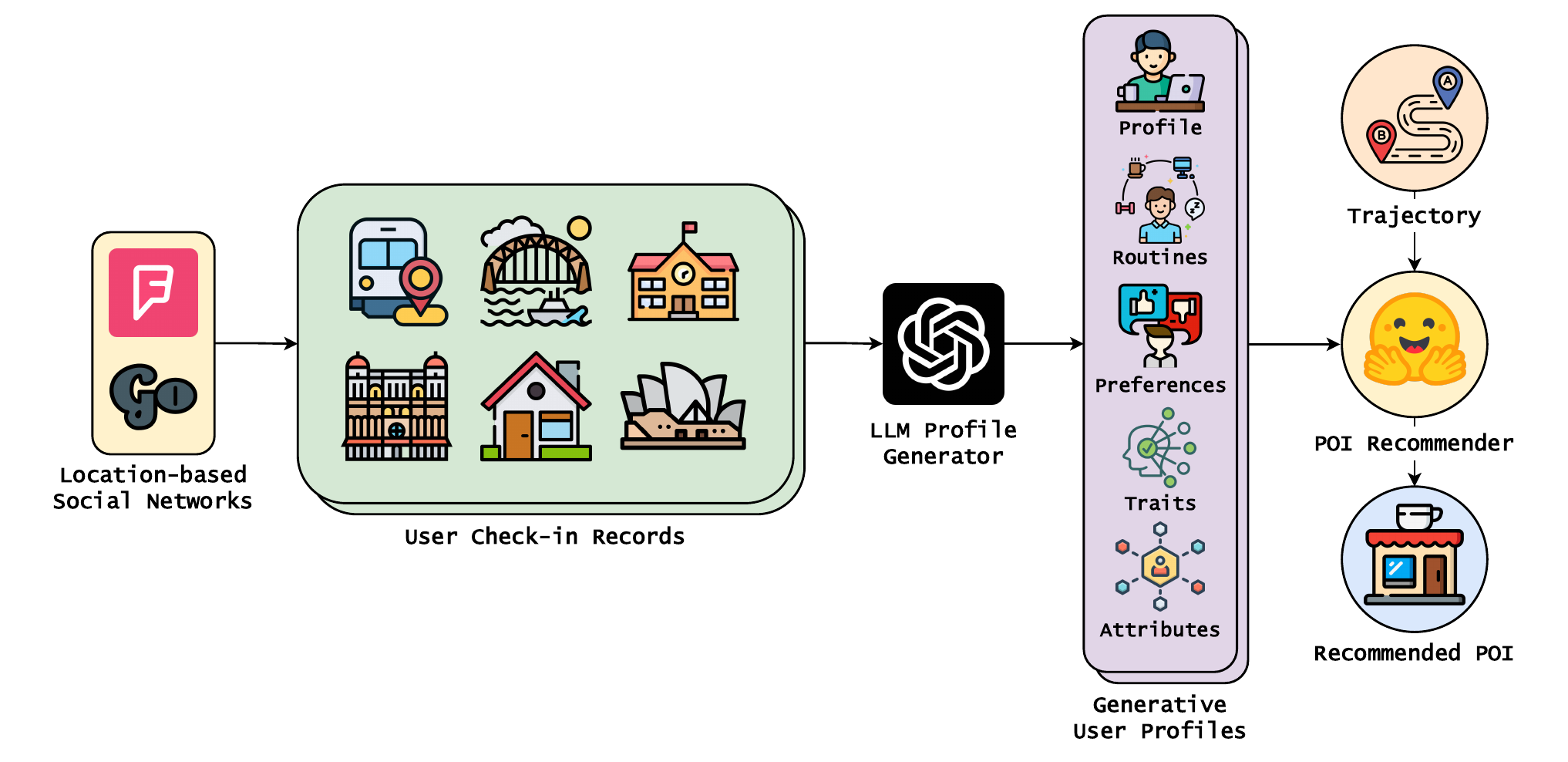}
  \caption{Our proposed framework of leveraging user check-in records as natural language user profiles as system prompts to large language models for supervised fine-tuning.}
  \Description{Our proposed framework of leveraging user check-in records as natural language user profiles as system prompts to large language models for supervised fine-tuning.}
  \label{fig:main-pipeline}
\end{figure*}

\subsection{Natural Language User Profile for Recommender Systems}
\label{sec:nl-up-rec-sys}

The use of natural language (NL) for user profiles in recommender systems marks a shift from traditional dense vector representations for user embeddings, addressing their limitations in transparency, interpretability, and scrutability \cite{radlinski2022natural}. This shift is driven by the rapid adoption of large language models (LLMs), which excel at generating human-readable text, enabling the automated creation of NL user profiles.

\paragraph{Advantages of NL User Profiles}
\citet{radlinski2022natural} highlighted key advantages of NL user profiles over vector representations. NL user profiles (1) enhance transparency by allowing users to see their preferences, (2) improve interpretability by making preferences understandable, and (3) offer scrutability to modify their profiles. 

NL profiles also help mitigate the cold-start problem, i.e. users who have not interacted much with a system. For example, new users can bootstrap their profiles by providing a concise textual description of their interests. Recommender systems can then leverage similar past profiles to infer similar patterns to generate recommendations \cite{radlinski2022natural}. Furthermore, NL profiles enable users to dynamically steer recommendations as changes to their NL profile are immediately reflected in the recommendations they receive, providing greater control and trust \cite{10.1145/1040830.1040870} in the system \cite{ramos2024natural}.

In terms of generalization, \citet{ramos2024natural} and \citet{zhou2024language} have demonstrated that NL profiles are adaptable across domains including movie, TV show, and hotel reviews, proving effective on recommender tasks such as pairwise preference, pairwise choice, and rating prediction. This showcases the flexibility of NL in capturing a wide range of user preferences, from specific attributes like favorite genres, user expertise, and broader behavioral patterns.

Given these advantages, our method leverages LLMs to generate NL user profiles for point of interest (POI) prediction tasks. These profiles need to encompass not only the user’s basic information but also capture their routines, preferences, and behavioral patterns to enhance the accuracy of POI recommendations. Our primary user behavior data source will be user check-in records from location-based social networks (LBSNs). We hypothesize that this will uncover users' patterns and routines, along with other relevant details to enhance the accuracy of POI recommendations. Unlike the other data sources mentioned, check-in records are unlabelled, exploiting large volumes of mobility data from online LBSNs that have been extensively collected and studied \cite{6844862,DBLP:journals/jnca/YangZCQ15,10.1145/2814575,10.1145/2020408.2020579,feng2019dplink}.

\paragraph{Challenges in NL User Profile Generation}
Despite the advantages, generating and optimizing NL profiles remains challenging. Profiles must be concise yet comprehensive, balancing short- and long-term preferences without overwhelming them with excessive information. Additionally, scaling NL profiles in large systems while avoiding biases in LLM-generated content remains difficult. Similarly, the evaluation of these profiles, in terms of fluency, coherence, and correctness, also lacks standardized metrics, complicating comparisons between methods \cite{radlinski2022natural,zhou2024language}. \cite{ramos2024natural} for example, conducted a qualitative case study where the profiles were evaluated by humans. Further optimizations include reinforcement learning algorithms, such as \verb|REBEL| \cite{gao2024rebel} which iteratively refine NL profile generation by using recommendation quality metrics and employing a reward model.

Several approaches have emerged, from early template-based methods \cite{balog2019transparent}, to more advanced LLM-driven techniques \cite{NEURIPS2020_1457c0d6}. Hybrid methods, such as those combining template-based models with LLMs, have been employed to generate more nuanced and contextually relevant profiles, leveraging sentiment analysis and feature extraction \cite{ramos2024natural}. Our method adopts a similar hybrid approach by grounding our NL profiles on user check-in records while also integrating robust personality assessment and behavioral theories such as the Theory of Planned Behavior \cite{ajzen1991theory,shao2024chainofplannedbehaviourworkflowelicitsfewshot}, Big Five Inventory (BFI) traits \cite{briggs1992assessing,de2000big,goldberg2013alternative,jiang2023personallm}, and user attributes \cite{chen2024designing}. This combination enhances the accuracy of the generated NL profiles by capturing the user’s underlying desires, intentions, and behavioral patterns, which can lead to more accurate POI predictions.

\section{Problem Definition}

We adopted the same problem definition as LLM4POI \cite{li2024large}, converting sequences of user check-in records from LBSNs into prompts and treating next POI prediction as a supervised question-answering task. In their setup, each user check-in record is represented as a tuple $q = (u, p, c, t)$, where $u \in U$ with $U$ denoting the set of all users. The variable $p$ refers to a point of interest (POI) from the set $P = \left\{p_1, p_2, \dots, p_M  \right\}$, where $M$ is the number of unique POIs. Additionally, $c$ denotes the POI’s category name, providing the LLM with extra semantic context about the POI, and $t$ is the timestamp of the check-in. We excluded the geometric coordinates (latitude, longitude) of the POIs, as they are irrelevant to the LLM \cite{li2024large}.

For each user $u$, trajectories were formed by grouping their check-in records into sequences within a specific time interval $\Delta t$. A trajectory $T_i^u(t)$ up to timestamp $t$ for user $u$ consists of a sequence of check-in:

\begin{equation*}
    T_i^u(t) = \{(p_1, c_1, t_1), \dots, (p_k, c_k, t_k)\},
\end{equation*}
where $t_1 < t_2 < \dots < t_k = t$ and $t_k - t_1 \leq \Delta t$. During training, the LLM is provided with a set of $L$ historical trajectories $\mathcal{T}_u = \{T_1^u, T_2^u, \dots, T_L^u\}$ for each user $u$, with the objective of predicting the next POI ID $p_{k+1}$, where the user will check in at the next timestamp $t_{k+1}$, given a current trajectory ${T^\prime}_i^u(t)$. The current trajectory refers to the list of POIs visited by the user prior to $t_{k+1}$ within the time interval $\Delta t$, while historical trajectories encompass previous trajectories occurring before the current time interval.

\section{Methodology}

Based on the findings of \cite{ramos2024natural} and building upon LLM4POI \cite{li2024large}, we hypothesize that NL user profiles can (1) replace long historical trajectories with a concise system prompt, (2) capture a user’s preferences and routines based on past check-ins, and (3) enable the LLM to learn from similar profiles to address the cold-start problem. Together, these capabilities aim to improve the efficiency and accuracy of next POI recommendation.

The overall framework we propose is illustrated in Figure \ref{fig:main-pipeline}. First, we generate NL user profiles from user check-in records, which are then used as system prompts for the LLM during the second stage of supervised fine-tuning (SFT). Unlike LLM4POI, our approach only used the user’s current trajectory as input, disregarding all historical trajectories from both the current and other users.

\subsection{Natural Language User Profile Generation}
\label{sec:nl-up-generation}

\paragraph{Generation} We describe our proposed NL user profile generation framework as follows. We utilized large-scale, unsupervised user check-in records from location-based social networks (LBSNs), enriched with semantic information (POI category name), as our primary source of user behavior data. GPT-4o Mini\footnote{We used gpt-4o-mini-2024-07-18 via the OpenAI API.} was used as the large language model (LLM) to generate NL user profiles. To ground our NL user profile generation, we propose the integration of three components:

\begin{itemize}

\item \textbf{Theory of Planned Behaviour} \cite{ajzen1991theory}. As demonstrated by \citet{shao2024chainofplannedbehaviourworkflowelicitsfewshot}, LLMs are capable of capturing a user’s attitude (preferences), subjective norms (routines), and perceived behavioral control (perceived likelihood of intentions based on profile, routine, and behavior history) from user activity logs/intention sequences. This improves the next activity and POI prediction. In our work, we focus on predicting user preferences and routines, such as frequently visited POIs and preferred POI categories.

\item \textbf{Big Five Inventory (BFI) personality traits} \cite{briggs1992assessing,de2000big,goldberg2013alternative}. \citet{jiang2023personallm} showed that pre-trained LLMs could generate content steered by the BFI personality traits they are asked to emulate, effectively emulating a specific user’s behavior and preferences. The BFI personality traits include: (1) extroverted/introverted, (2) agreeable/antagonistic, (3) conscientious/unconscientious, (4) neurotic/emotionally stable, and (5) open/closed to experience.

\item \textbf{User Attributes}. \citet{chen2024designing} showed that LLMs can reliably predict user attributes such as age, gender, educational background, and socioeconomic level. These attributes were intentionally selected as additional prompt information to the LLM as they are pivotal in shaping and making critical real-world decisions. Following TalkTuner \cite{chen2024designing}, we categorized the four attributes as follows: age (child: <13, adolescent: 13-17, adult: 18-64, older adult: >64); gender (male/female); educational background (some schooling, high school, college \& beyond); and socioeconomic status (lower, middle, upper).

\end{itemize}

Following the approach of \cite{li2024large} and using their prompt formatting structure, user check-in records and trajectories were first transformed into prompts. Trajectory prompting \cite{radford2021learning,xue2022leveraging,zhou2022learning} allows the conversion of heterogeneous data from LBSNs into natural language sentences that LLMs can understand. These trajectory prompts were then concatenated and used to generate NL user profiles, as outlined in Table \ref{tab:user-profile-prompting}. In addition to generating a 200-word NL user profile, the LLM was tasked with predicting the 3 components above. To facilitate parsing, the output is returned in JSON format. 

\subsubsection{User Profile Validation and Challenges}


A key challenge in user profile generation is validation, as extensively discussed in \S \ref{sec:nl-up-rec-sys}. Another challenge we encountered was LLM hallucinations, where it failed to follow the instructed JSON schema, leading to inconsistent parsing due to changed JSON key names. However, this issue affected fewer than five users and was easily resolved by re-running the prompt. To validate our generated profiles, we first ensured that predicted personality traits were valid BFI categories and that user attributes matched predefined classes.

Additionally, we employed the LLM-as-a-Judge method \cite{zheng2023judging}, where LLMs assessed the quality of generated profiles. Following the qualitative evaluation approach proposed by \cite{ramos2024natural}, we adapted their fluency, informativeness, conciseness, and relevance metrics to assess our NL profiles. LLMs were prompted using the metrics shown in Table~\ref{tab:user-profile-validation}, and their evaluation results are shown in Table~\ref{tab:user-profile-validation-results}. To mitigate bias from a single model’s judgment, we conducted evaluations using two LLMs: GPT-4o Mini and Gemini 1.5 Flash. Notably, conciseness emerged as a consistent weakness across both judges, mirroring observations from \cite{ramos2024natural}, where human annotators similarly flagged verbosity in LLM-generated profiles. However, on the other metrics, our profiles were rated favorably, indicating that despite verbosity, they remain fluent, informative, and relevant.

To further investigate potential biases in our generated NL user profiles, we analyzed their distributions across three datasets: NYC, CA, and TKY. Table~\ref{tab:user-profile-traits} presents the distribution of predicted BFI traits, while Table~\ref{tab:user-profile-attributes} summarizes the user attribute distributions. These statistics highlight the diversity of inferred user characteristics within and across datasets, providing insight into the diversity and variability of the generated profiles.

\begin{table}[tbp]
\small
\centering
\caption{Natural language (NL) user profile generation prompt and check-in record trajectory prompt used for user profile generation. The teal text instructs the LLM to generate a 200-word user profile, orange for predicting preferences and routines, red for predicting BFI traits, and blue for predicting user attributes.}
\label{tab:user-profile-prompting}
\begin{tabular}{p{1.5cm}p{6cm}}
\toprule
\textbf{prompt} & Given the following check-ins of user [user id], generate \textcolor{teal}{a 200-word user profile summary} to be used as a system prompt to another LLM that simulates this person's behavior, \textcolor{orange}{preferences, routines}, hobbies, schedule, etc. Predict this user's \textcolor{red}{Big Five Inventory traits: (1) extroverted / introverted, (2) agreeable / antagonistic, (3) conscientious / unconscientious, (4) neurotic / emotionally stable, (5) open / closed to experience.} \textcolor{blue}{Also predict their age: child (<13) / adolescent (13-17) / adult (18-64) / older adult (>64); their gender: male / female; their educational background: some schooling / high school / college \& beyond; and their socioeconomic level: lower / middle / upper.} Also include any patterns that is observed and POI IDs of important places that might be visited in the future in the user profile summary. This system prompt will be used to make future check-in predictions. Return your response in JSON format: \{\textcolor{red}{"traits": [trait1, trait2, trait3, trait4, trait5]}, \textcolor{blue}{"attributes": [age, gender, edu, socioeco]}, \textcolor{orange}{"preferences": [preference1, preference2, preference3, ...], "routines": [routine1, routine2, routine3, ...]}, \textcolor{teal}{"user\_profile": "200-word user profile summary"}\} [check-in records]. \\
\midrule
\textbf{check-in record} & At [time], user [user id] visited POI id [poi id] which is a/an [poi category name] with category id [poi category id]. \\
\bottomrule
\end{tabular}
\end{table}

\begin{table}[htbp]
\small
\centering
\caption{User profile validation prompt provided to LLM judge for evaluation, using metrics from \cite{ramos2024natural}.}
\label{tab:user-profile-validation}
\begin{tabular}{p{8cm}}
\toprule
Given the following natural language (NL) user profile [user profile], you are to assess the user profile based on four criteria: \\
- \textbf{Fluency}: Is the NL profile both syntactically and semantically correct? \\
- \textbf{Informativeness}: Does the NL profile provide important information for a user profile? \\
- \textbf{Conciseness}: Is the NL profile written in a concise manner? \\
- \textbf{Relevance}: Given the list of reviews, is the NL profile relevant to the user? \\
Return your response in the JSON format: \{"fluency": 0/1, "informativeness": 0/1, "conciseness": 0/1, "relevance": 0/1\} \\
\bottomrule
\end{tabular}
\end{table}

\begin{table}[htbp]
\small
\centering
\caption{Evaluation results of LLM-judged user profile validation using metrics from \cite{ramos2024natural}, conducted on generated profiles from users in New York City, Tokyo, and California.}
\label{tab:user-profile-validation-results}
\begin{tabular}{lccc}
\toprule
\textbf{Metric} & \textbf{NYC} & \textbf{TKY} & 
\textbf{CA} \\
\midrule
\multicolumn{4}{l}{\textit{Judge: GPT 4-o Mini}} \\
\midrule
Fluency         & 100.00       & 100.00      & 100.00       \\
Informativeness & 100.00       & 100.00      & 100.00       \\
Conciseness     & 43.17        & 40.60       & 37.26        \\
Relevance       & 100.00       & 100.00      & 100.00    \\
\midrule
\multicolumn{4}{l}{\textit{Judge: Gemini 1.5 Flash}} \\
\midrule
Fluency         & 100.00       & 100.00      & 100.00       \\
Informativeness & 100.00       & 100.00      & 100.00       \\
Conciseness     & 65.81        & 63.48       & 57.68        \\
Relevance       & 100.00       & 100.00      & 100.00    \\
\bottomrule
\end{tabular}
\end{table}

\subsection{Trajectory Prompting with Natural Language User Profiles}

Likewise, we followed a similar prompt structure to LLM4POI, with modifications to include NL user profiles and the use of Llama 2’s \cite{touvron2023llama2openfoundation} chat prompt template, as shown in Table \ref{tab:sft-prompt}. Each prompt contains the user’s NL profile, the current trajectory block, an instruction block, and the corresponding target block. With LLM4POI’s setup, heterogenous LBSN data---such as user IDs, timestamps, and POI category names---are converted into sentences that capitalize on the LLM’s strong semantic understanding capabilities.

The NL user profile generated using the method described in \S \ref{sec:nl-up-generation} serves as the system prompt for the LLM, providing a description of the user that the model is intended to simulate. Given that the output is formatted in JSON, with each component represented as a different key, it can be easily parsed and integrated into the input prompt. This system prompt informs the LLM of the user’s attributes, including age, gender, educational background, socioeconomic level, BFI traits, preferences, and routines. Consequently, the LLM's POI prediction output can leverage the insights provided by the NL user profile. Additionally, the LLM can learn trajectory patterns and routines from users with similar NL profiles, thereby enhancing its predictive capabilities. 

The current trajectory block consists of all check-in records from the current session, except for the latest record, which is placed in the target block. Unlike LLM4POI, we excluded historical check-in records from the user and other users to isolate the impact of our proposed NL user profiles and minimize the input context length.\footnote{Our method only implicitly incorporates historical information into its user profile, unlike LLM4POI. The latter has access to other similar trajectories, whereas the former can only summarize frequent or long-term trajectories into a concise user profile. LLM4POI also incorporates intra-user social data, while ours does not.} The range of POI IDs specified in the instruction block helps guide the LLM in generating a predicted POI ID, ensuring it falls within the set of POIs $P$. The target block, which includes the next timestamp, user ID, and next POI ID, serves as the generation targets/labels and is excluded from the input during the prediction process.

\begin{table}[tbp]
\small
\centering
\caption{Distribution of predicted BFI traits across the New York City, Tokyo, and California datasets.}
\label{tab:user-profile-traits}
\begin{tabular}{lccc}
\toprule
\textbf{Trait} & \textbf{NYC} & \textbf{TKY} & \textbf{CA} \\
 \midrule
Extroverted & 736 & 1476 & 3526 \\
Introverted & 311 & 805 & 430 \\
 \midrule
Agreeable & 1047 & 2281 & 3956 \\
Antagonistic & 0 & 0 & 0 \\
 \midrule
Conscientious & 1047 & 2281 & 3956 \\
Unconscientious & 0 & 0 & 0 \\
 \midrule
Neurotic & 41 & 99 & 268 \\
Emotionally Stable & 1006 & 2182 & 3688 \\
 \midrule
Open to Experience & 1043 & 2279 & 3947 \\
Closed to Experience & 4 & 2 & 9 \\
\bottomrule
\end{tabular}
\end{table}

\begin{table}[tbp]
\small
\centering
\caption{Distribution of predicted user attributes across the New York City, Tokyo, and California datasets.}
\label{tab:user-profile-attributes}
\begin{tabular}{llccc}
\toprule
\textbf{Attribute} &  & \textbf{NYC} & \textbf{TKY} & \textbf{CA} \\
 \midrule
\multirow{4}{*}{Age} & Child & 1 & 2 & 14 \\
 & Adolescent & 72 & 274 & 330 \\
 & Adult & 974 & 2005 & 3612 \\
 & Older Adult & 0 & 0 & 0 \\
 \midrule
\multirow{2}{*}{Gender} & Male & 774 & 1881 & 2911 \\
 & Female & 273 & 400 & 1045 \\
 \midrule
\multirow{3}{*}{Education} & Some Schooling & 43 & 192 & 190 \\
 & High School & 55 & 165 & 176 \\
 & College \& Beyond & 949 & 1924 & 3590 \\
 \midrule
\multirow{3}{*}{Socioeconomic} & Lower & 3 & 3 & 7 \\
 & Middle & 1028 & 2240 & 3651 \\
 & Upper & 16 & 38 & 298 \\
 \bottomrule
\end{tabular}
\end{table}

\begin{table}[tbp]
\small
\centering
\caption{Structure of NL user profile as system prompt, input prompt, and check-in record for supervised fine-tuning based on Llama 2's chat prompt template. The teal text represents the system prompt, red for the current trajectory prompt, orange for the instruction prompt, and blue for the target block.}
\label{tab:sft-prompt}
\begin{tabular}{p{1.3cm}p{6.5cm}}
\toprule
\textbf{system prompt} & \textcolor{teal}{You are user [user id] and your basic information is as follows: Age: [age]; Gender: [gender]; Education: [education]; SocioEco: [socioeco]. You have the following traits: [trait1], [trait2], [trait3], [trait4], [trait5]. You have the following preferences: [preference1], [...], [preferenceN]. You have the following routines: [routine1], [...], [routineM]. [user profile]}\\
\midrule
\textbf{prompt} & <s>[INST] \textcolor{teal}{\textnormal{<}\textnormal{<}SYS\textnormal{>}\textnormal{>} [system prompt] \textnormal{<}\textnormal{<}/SYS\textnormal{>}\textnormal{>}} \textcolor{red}{The following is a trajectory of user [user id]: [check-in records].} \textcolor{orange}{Given the data, at time [time], which POI id will user [user id] visit? Note that POI id is an integer in the range from 0 to [POI id range].} [/INST] \textcolor{blue}{At [time], user [user id] will visit POI id [poi id].} </s>\\
\midrule
\textbf{check-in record} & At [time], user [user id] visited POI id [poi id] which is a/an [poi category name] with category id [poi category id].\\
\bottomrule
\end{tabular}
\end{table}

\subsection{Supervised Fine-tuning}

Moreover, we replicated a similar compute- and memory-efficient supervised fine-tuning (SFT) setup as LLM4POI \cite{li2024large} to ensure a fair comparison with their approach as our baseline.

\paragraph{Parameter-Efficient Fine-tuning (PEFT)} Like LLM4POI, we applied LoRA \cite{hu2021lora} as the parameter-efficient fine-tuning (PEFT) technique used during training, which involves freezing all parameters in the pre-trained LLM and only training the newly injected low-rank decomposition matrices. These matrices are subsequently merged into the pre-trained LLM during the inference process. In their setup, LoRA matrices of rank 8 were injected into the attention layers (query, key, value, and output projection layers). This significantly reduces the GPU memory requirements when fine-tuning an LLM while maintaining similar performances to full fine-tuning.

\paragraph{Quantized Training} To further reduce GPU memory usage during the fine-tuning process, we applied the same quantization technique used by LLM4POI. Namely, we applied QLoRA (Quantized LoRA) \cite{dettmers2023qloraefficientfinetuningquantized} which combines LoRA with a novel 4-bit NormalFloat (NF4) data type for storing model weights, and applies double quantization. All forward and backward passes are still conducted in the BrainFloat16 (BF16) data type.

\paragraph{Efficient Training Kernels} As the trajectory prompts can be quite long after tokenization, we require a long context length of at least 16,384 tokens to fit our input prompt without truncation. LLM4POI utilized FlashAttention-2 (FA2) \cite{dao2022flashattentionfastmemoryefficientexact,dao2023flashattention2fasterattentionbetter} in their training implementation to reduce the GPU memory usage and accommodate their 32,768-token sequence length. We similarly leveraged FA2 as our efficient scaled dot product attention (SDPA) kernel implementation, as well as the newly introduced Liger Kernel \cite{hsu2024ligerkernelefficienttriton} that increases training throughput and reduces memory usage. Like FA2, Liger Kernel introduced efficient GPU kernel implementations for other common LLM layers and operations such as RoPE \cite{su2023roformerenhancedtransformerrotary}, RMSNorm \cite{zhang2019rootmeansquarelayer}, SwiGLU \cite{shazeer2020glu}, and cross-entropy loss.

\section{Experiments}

\subsection{Setup}

\begin{figure}
    \centering
    \hfill
    \begin{subfigure}[t]{0.225\textwidth}
        \centering
        \includegraphics[width=\textwidth]{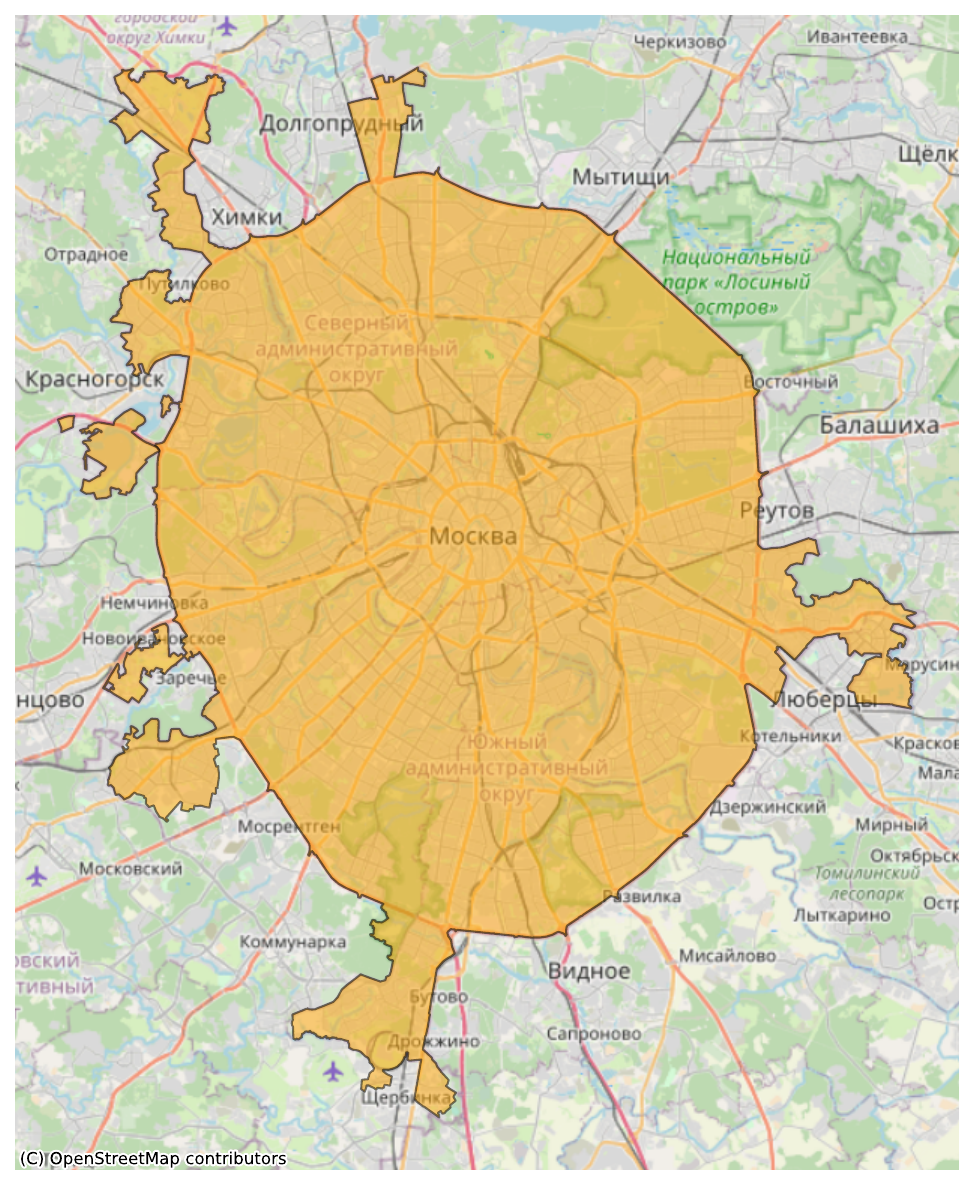}
        \caption{Moscow, Russia}
    \end{subfigure}
    \hfill
    \begin{subfigure}[t]{0.18\textwidth}
        \centering
        \includegraphics[width=\textwidth]{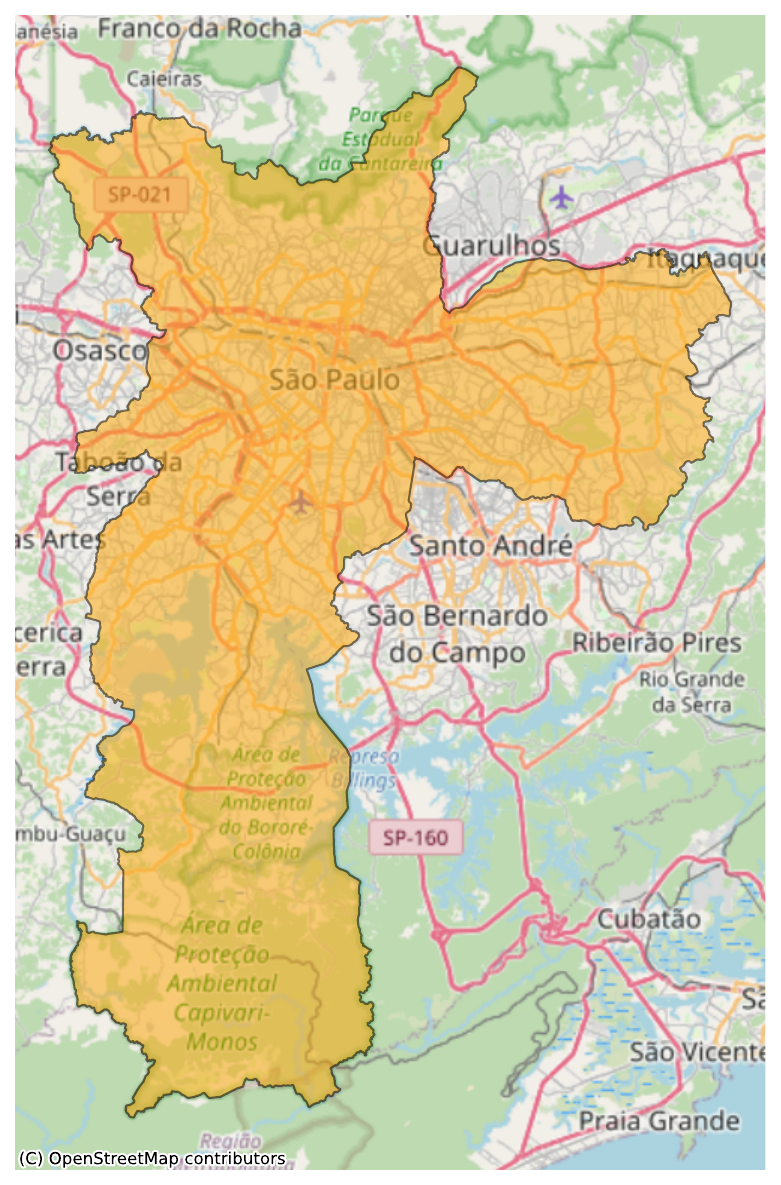}
        \caption{São Paulo, Brazil}
    \end{subfigure}
    \hfill
    \caption{Maps of the recreated Moscow and São Paulo subsets from the Global-scale Check-in Dataset.}
    \Description{Maps of the recreated Moscow and São Paulo subsets from the Global-scale Check-in Dataset.}
    \label{fig:geom-maps}
\end{figure}

\subsubsection{Datasets}

As we aimed to compare our approach with that of LLM4POI, we used the same public datasets they used and employed their pre-processing code\footnote{\url{https://github.com/neolifer/LLM4POI}}. Specifically, we used the New York City (NYC) and Tokyo (TKY) Check-in Dataset \cite{6844862}, which contains user check-ins from 12 April 2012 to 16 February 2013 from the popular LBSN, Foursquare. They also used the Gowalla-CA dataset \cite{yan2023spatio}, a subset of the original Gowalla dataset \cite{10.1145/2020408.2020579}, and filtered for check-in records from California and Nevada using geometric coordinates. Following their procedures, the filtered and pre-processed records are divided into training, validation, and test sets, using the first 80\% for training, the middle 10\% for validation, and the final 10\% for testing. Only users and POIs that appear in the training set are kept in the validation and testing sets.

Furthermore, we recreated Moscow and São Paulo subsets of the Global-scale Check-in Dataset \cite{yang2019revisiting,9099985}, following the pre-processing methods outlined by \cite{feng2024agentmove}. Unlike their approach, we used actual geometric boundaries of the cities obtained from OpenStreetMap\footnote{\url{https://www.openstreetmap.org/}} and included only the POIs located within those boundaries. Trajectories were grouped into 72-hour sessions, with users having fewer than 5 sessions and sessions with fewer than 4 check-ins removed. The data was split into 70\% for training, 10\% for validation, and 20\% for testing. Only users with between 3 and 50 sessions were included in the testing set. Like \cite{feng2024agentmove}, we limited the test set to 200 samples during evaluation due to cost constraints, but fine-tuned our models on the entire training set. All of the datasets we used are publicly available and anonymous. The maps of geometric boundaries of the two cities are shown in Figure \ref{fig:geom-maps}. 


\subsubsection{Baselines}

We compared our approach with the three kinds of baselines: supervised fine-tuned (SFT) LLMs, LLM-based in-context learning methods, and graph-based models.

\begin{itemize}
    \item LLM4POI \cite{li2024large}: Main baseline, leveraging LLM for next POI recommendation by utilizing similar historical trajectories from current users and other users. Frames next POI prediction task as question-answer SFT pairs.
    \item LLM-Mob \cite{wang2023would}: Employs in-context learning to enhance POI recommendations by providing historical and contextual user trajectories. Uses extensive prompting and zero-shot reasoning capabilities of LLMs.
    \item LLM-ZS \cite{beneduce2024large}: A simplified version of LLM-Mob, also utilizes historical and contextual trajectories, simpler prompts, and zero-shot reasoning. Compares zero-shot, one-shot, and few-shot prompting predictions.
    \item GETNext \cite{yang2022getnext}: Graph-enhanced transformer model that incorporates a global trajectory flow map to leverage collaborative signals from other users and POI embeddings.
    \item STHGCN \cite{yan2023spatio}: A spatio-temporal framework that models higher-order dependencies in time and space, using a hypergraph transformer to model cross-user relations.
\end{itemize}

\subsubsection{Evaluation}


We evaluated our model using Accuracy@1, aligning with the evaluation protocol of LLM4POI to ensure fair and consistent comparison. LLM4POI motivates the use of Accuracy@1 by framing \textit{next} POI recommendation as a question-answering task, where only a single, contextually appropriate destination is returned, reflecting realistic constraints such as time-sensitive travel scenarios where only one top recommendation is allowed.

\subsubsection{Models}
\label{llm4poi-asterisk}

To ensure a fair comparison against the reported results of LLM4POI \cite{li2024large}, we used the same pre-trained LLM, Llama-2-7b-longlora-32k \cite{chen2024longloraefficientfinetuninglongcontext}, for our main comparative experiments. In their setup, the model without historical trajectories is denoted as LLM4POI*, while the model trained with historical trajectories from both the current user and other users is called LLM4POI. We denote our models with the prefix GenUP (generative user profilers). Moreover, we fine-tuned more recent variants of the LLaMA family of models \cite{touvron2023llamaopenefficientfoundation,dubey2024llama3herdmodels}, including Llama 3.1 8B and Llama 3.2 1B, to further benchmark our method against newer architectures and evaluate their performance on the next POI prediction task.

\subsubsection{Implementation}

Our fine-tuning implementation follows the setup described in \cite{li2024large}. We used a linearly decaying learning rate scheduler for all fine-tuning runs, with 20 warmup steps and no weight decay. Since we excluded historical trajectories from our prompts, we used a context length of 16,384 tokens and trained for 3 epochs. The Transformer Reinforcement Learning (TRL) library \cite{vonwerra2022trl,wolf2020huggingfacestransformersstateoftheartnatural} was used throughout our SFT experiments, which were conducted on NVIDIA H100 GPUs.

\subsection{Main Results}

We present our results in comparison to the baseline LLM4POI approach in Table \ref{tab:results-1}. Additionally, we included their reported results for sequence-based and graph-based models to emphasize the performance of our proposed method against those that leverage additional historical trajectories of current users, as well as collaborative and intra-user social data. 

The evaluation results from the NYC, TKY, and CA datasets demonstrate the effectiveness of our proposed method, which consistently outperforms LLM4POI*, both of which utilize the same Llama 2 base LLM. Specifically, GenUP-Llama-2-7B achieves higher accuracy in NYC (0.2575 vs. 0.2356), TKY (0.1699 vs. 0.1517), and CA (0.1094 vs. 0.1016). These findings indicate that incorporating natural language user profiles as system prompts improves performance, even when using the same pre-trained LLM. Moreover, the larger GenUP-Llama-3.1-8B model, further increases performance, achieving the highest accuracy in all three cities (NYC: 0.2582, TKY: 0.2127, CA: 0.1339). Notably, this model demonstrated competitive results compared to GETNext which exploits collaborative signals from cross-user data. Similarly, the smaller GenUP-Llama-3.2-1B model also consistently surpasses the LLM4POI* baseline, showing an improvement in accuracy despite using a smaller yet more powerful pre-trained base LLM.

We also compare the results of our proposed method with LLM-based baselines that rely on historical trajectories and employ in-context learning (ICL) rather than supervised fine-tuning (SFT). To ensure a fair comparison, we re-implemented LLM-Mob and LLM-ZS using Llama 3.1 8B Instruct to match the capabilities of base models. We present the evaluation results in Table~\ref{tab:results-2}. Our models consistently outperform LLM-Mob and LLM-ZS on both the Moscow and São Paulo datasets, despite those baselines leveraging ICL with historical trajectories as input. In contrast, GenUP uses compact NL user profiles to implicitly learn historical trajectories without relying on long context windows.

\begin{table}[tbp]
\centering
\caption{Evaluation results of our proposed method incorporating natural language user profiles as system prompts for LLM-based next POI prediction on New York City, Tokyo, and California datasets. The highest score on each dataset is highlighted in \textbf{bold}, while the second highest is \underline{underlined}.}
\label{tab:results-1}
\begin{tabular}{lccc}
\toprule
\textbf{Model} & \textbf{NYC} & \textbf{TKY} & \textbf{CA} \\
\midrule
\multicolumn{4}{l}{\textit{Without historical and intra-user social data}} \\
\midrule
LLM4POI* & 0.2356 & 0.1517 & 0.1016 \\
GenUP-Llama-2-7B & \underline{0.2575} & 0.1699 & 0.1094 \\
GenUP-Llama-3.1-8B & \textbf{0.2582} & \textbf{0.2127} & \textbf{0.1339} \\
GenUP-Llama-3.2-1B & 0.2484 & \underline{0.1851} & \underline{0.1267} \\
\midrule
\multicolumn{4}{l}{\textit{With historical and intra-user social data}} \\
\midrule
GETNext & 0.2435 & 0.2254 & 0.1357 \\
STHGCN & 0.2734 & 0.2950 & 0.1730 \\
LLM4POI & 0.3372 & 0.3035 & 0.2065 \\
\bottomrule
\end{tabular}
\end{table}

\begin{table}[tbp]
\small
\centering
\caption{Evaluation results of our method compared to in-context learning LLM-based methods for next POI prediction on Moscow and São Paulo datasets. The highest score on each dataset is highlighted in \textbf{bold}, while the second highest is \underline{underlined}.}
\label{tab:results-2}
\begin{tabular}{llccc}
\toprule
\textbf{Model} & \textbf{Base Model} & \textbf{Moscow} & \textbf{São Paulo} \\
\midrule
\multicolumn{4}{l}{\textit{In-context learning with historical data}} \\
\midrule
LLM-ZS & Llama 3.1 8B Inst. & 0.115 & \underline{0.130} \\
LLM-Mob & Llama 3.1 8B Inst. & 0.145 & 0.120 \\
\midrule
\multicolumn{4}{l}{\textit{Supervised fine-tuning}} \\
\midrule
GenUP-Llama-2-7B & Llama 2 7B & \textbf{0.180} & \textbf{0.205} \\
GenUP-Llama-3.1-8B & Llama 3.1 8B & \underline{0.170} & \textbf{0.205} \\
GenUP-Llama-3.2-1B & Llama 3.2 1B & \textbf{0.180} & \textbf{0.205} \\
\bottomrule
\end{tabular}
\end{table}

\subsection{Analysis}

\subsubsection{Impact of Natural Language User Profile Components}
\label{sec:impact-components}

We investigated the impact of different components used to create our NL user profile, with the results shown in Table \ref{tab:ablations}. In particular, we tested the following combination of components to include in the system prompt during training and inference:

\begin{itemize}
    \item \textbf{No user profile:} Baseline method of LLM4POI* with no system prompts.
    \item \textbf{User Profile:} A short, 200-word user profile.
    \item \textbf{User Profile + Routines \& Preferences:} The user profile along with the predicted routines and preferences based on the Theory of Planned Behavior.
    \item \textbf{User Profile + Routines \& Preferences + Attributes:} The user profile, predicted routines and preferences, and predicted user attributes (age, gender, educational background, socioeconomic status).
    \item \textbf{User Profile + Routines \& Preferences + Attributes + BFI Traits:} The full set, including the user profile, predicted routines and preferences, predicted user attributes, and predicted Big Five Inventory traits.
\end{itemize}

The results demonstrate that adding a simple, 200-word user profile generated from check-in records already improves the baseline performance of LLM4POI*, indicating that our approach of adding a system prompt enables the LLM to better emulate user behavior during POI prediction. However, not all predicted components lead to improved performance. For example, adding routines, preferences, and BFI traits does not impact the model’s predictive accuracy. On the other hand, the inclusion of predicted user attributes provides a slight improvement, suggesting that these attributes influence decision-making processes and user behavior, as proposed by \cite{chen2024designing}.

\begin{table}[tbp]
\small
\centering
\caption{Ablation study of the different natural language user profile components as system prompts on the NYC dataset. The highest score is highlighted in \textbf{bold}.}
\label{tab:ablations}
\begin{tabular}{llc}
\toprule
\textbf{System Prompt}       & \textbf{Model}  & \textbf{NYC}    \\
\midrule
No user profile          & LLM4POI*        & 0.2356          \\
200-word User Profile              & GenUP-Llama2-7B & 0.2568          \\
+ Routines \& Preferences & GenUP-Llama2-7B & 0.2568          \\
+ User Attributes              & GenUP-Llama2-7B & \textbf{0.2575} \\
+ BFI Traits              & GenUP-Llama2-7B & \textbf{0.2575} \\
\bottomrule
\end{tabular}
\end{table}

\subsubsection{User Cold-start Analysis}

Inspired by \cite{li2024large}, we also conducted a user cold-start analysis on our proposed GenUP models across all five datasets in our study. Following their assessment method, users were categorized into different activity levels based on the number of trajectories in the training set. The top 30\% of users were classified as active and the bottom 30\% as inactive.

The results, presented in Table \ref{tab:cold-start-analysis}, illustrate a clear trend where the POI prediction accuracy increases with higher user activity levels, which aligns with intuitive expectations. For instance, the GenUP-Llama-2-7B model shows a progression in accuracy on the São Paulo dataset from 0.1366 for inactive users, to 0.1504 for normal users, and 0.1940 for very active users. This pattern holds across most models and cities, except for the CA dataset, where the models are slightly more accurate for inactive users, consistent with the findings reported by LLM4POI \cite{li2024large}.

Furthermore, larger models such as GenUP-Llama-3.1-8B consistently achieve better performance in very active user groups. For example, on the TKY dataset, GenUP-Llama3.1-8B achieves an accuracy of 0.2688, compared to 0.2063 for GenUP-Llama-2-7B. These results suggest that larger, more sophisticated models benefit from richer data available for very active users, leading to substantial gains in prediction accuracy. However, the variability in performance for inactive users indicates that the advantages of larger models are diminished when user activity data is limited.

We also analyzed how different NL user profile components affect different user groups, with results shown in Table \ref{tab:cold-start-components}. Our findings indicate that their impact varies by user activity level. For inactive (cold-start) users, incorporating routines and preferences improved accuracy by +0.93\%, with user attributes providing another +0.93\% boost. However, while adding BFI traits reduced accuracy for inactive users -- as also observed in \S \ref{sec:impact-components} -- they enhanced the performance for highly active users.

\begin{table}[tbp]
\small
\centering
\caption{Impact of different NL user profile components on next POI recommendation accuracy for NYC users, categorized by activity level, using the GenUP-Llama2-7B model.}
\label{tab:cold-start-components}
\begin{tabular}{lccc}
\toprule
\textbf{System Prompt} & \textbf{Inactive} & \textbf{Normal} & \textbf{Very Active} \\
\midrule
200-word User Profile  & 0.2012  & 0.2737  & 0.2728  \\
+ Routines \& Preferences  & 0.2105  & 0.2774  & 0.2680  \\
+ User Attributes  & 0.2198  & 0.2664  & 0.2692  \\
+ BFI Traits  & 0.2105  & 0.2591  & 0.2752  \\
\bottomrule
\end{tabular}
\end{table}

\begin{table}[tbp]
\small
\centering
\caption{User cold-start analysis based on the number of trajectories on NYC, TKY, CA, Moscow and São Paulo datasets.}
\label{tab:cold-start-analysis}
\begin{tabular}{lccccc}
\toprule
\textbf{User Groups} & \textbf{NYC} & \textbf{TKY} & \textbf{CA} & \textbf{Moscow} & \textbf{São Paulo} \\
\midrule
\multicolumn{6}{l}{\textit{GenUP-Llama-2-7B}} \\
\midrule
Inactive             & 0.2105       & 0.1306       & 0.1091      & 0.1227          & 0.1366             \\
Normal               & 0.2591       & 0.1394       & 0.1089      & 0.1410          & 0.1504             \\
Very Active           & 0.2752       & 0.2063       & 0.1096      & 0.1748          & 0.1940             \\
\midrule
\multicolumn{6}{l}{\textit{GenUP-Llama-3.1-8B}} \\
\midrule
Inactive            & 0.1826       & 0.1486       & 0.1380      & 0.1180          & 0.1393             \\
Normal               & 0.2554       & 0.1695       & 0.1338      & 0.1464          & 0.1598             \\
Very Active          & 0.2884       & 0.2688       & 0.1324      & 0.1808          & 0.1944             \\
\midrule
\multicolumn{6}{l}{\textit{GenUP-Llama-3.2-1B}} \\
\midrule
Inactive             & 0.1764       & 0.1306       & 0.1316      & 0.1210          & 0.1429             \\
Normal                & 0.2664       & 0.1494       & 0.1223      & 0.1390          & 0.1530             \\
Very Active          & 0.2704       & 0.2321       & 0.1263      & 0.1793          & 0.1906  \\
\bottomrule
\end{tabular}
\end{table}

\subsubsection{Efficiency Comparison of Trajectory Similarity and User Profiling}

Trajectory similarity computation is crucial for overcoming the cold-start problem, as it helps predict inactive users’ preferences through past historical trajectories. Traditional methods often use methods like Dynamic Time Warping (DTW), to measure point-to-point similarity between trajectories. For two trajectories with an average length of k, the time complexity of a single DTW comparison is $O(k^2)$. When considering $L$ trajectories, the total complexity for comparing all pairs is $O(L^2 \times k^2)$. Additionally, sorting algorithms are typically required for ranking similarities, contributing an additional overhead of $O(L \log L)$, bringing the overall complexity to $O(L^2 \times k^2 + L \log L)$.

LLM4POI \cite{li2024large} proposed to enhance predictive accuracy by leveraging historical trajectories but faces challenges, including a long context length of 32,768 tokens needed to fit the entire input prompt. Although their trajectory embedding approach reduces the complexity of trajectory embedding inference to $O(L)$, the necessity of performing pairwise similarity comparisons among $L$ trajectories results in a time complexity of $O(L^2)$. With similarity sorting, the overall complexity becomes $O(L^2 +L \log L)$. While this method improves prediction accuracy, it still incurs substantial computational costs, particularly in scenarios with frequent updates, such as when new trajectories are available and when user preferences shift.

In contrast, our method replaces frequent trajectory similarity computations by utilizing user profiles that require only a single inference per user given their historical trajectories $\mathcal{T}_u$. Consequently, the time complexity for generating these profiles is only $O(|U|)$ where $|U|$ is the number of users and $|U| \ll L$.  These profiles can be easily updated periodically as new trajectories become available, ensuring they reflect evolving user behaviors without constant recalibration. An example is to employ a user profile embedding shift, where a new profile  $P^\prime$ replaces an existing profile $P$ if their similarity falls below a predefined threshold, indicating a significant behavioral shift. This strategy prioritizes meaningful long-term changes over fixed update intervals while allowing short-term adaptations when necessary. In practice, user profile generation with GPT-4o Mini averaged under 2 seconds per user and can be further accelerated with batched and parallel inference, making updates efficient even at scale. Notably, our method omits historical trajectories halving the context length of LLM4POI. This significantly reduces memory and compute requirements.


\begin{table}[tbp]
\small
\centering
\caption{Effect of short-term preference injection on predicted next POI category using GenUP-Llama-3.2-1B on the NYC test set. For each target POI category, we report the number of predicted visits before and after appending a short-term preference statement to the NL user profile.}
\label{tab:preference-shift}
\begin{tabular}{lcc}
\toprule
\textbf{Target Category} & \textbf{Original Profile} & \textbf{Added New Preference} \\
\midrule
Coffee Shop & 61 & 68 (+7) \\
Bar & 94 & 105 (+11) \\
Gym / Fitness Center & 102 & 123 (+21) \\
Subway & 53 & 65 (+12) \\
\bottomrule
\end{tabular}
\end{table}

\subsubsection{Preference Shift Adaptability}

To evaluate the responsiveness of GenUP to short-term user preference changes, we conducted an experiment on the NYC test set using the GenUP-Llama-3.2-1B model. Following the approach of prior work \cite{ramos2024natural}, we simulate short-term shifts in user preference by appending a new preference statement to the initial NL user profile of all users. We added the prompt \texttt{"Today, this user really wants to visit a \{POI category\} place."} to the NL user profile and repeated the evaluation process without retraining the model. We selected four commonly visited POI categories for evaluation: Coffee Shop, Bar, Gym/Fitness Center, and Subway Station. For each category, we measured the number of predicted visits to that POI category before and after injecting the preference prompt into the user's profile. The evaluation focuses on the relative change in predicted category frequencies, with the expectation that predictions will shift toward the injected preference.

Naturally, the degree of change depends on multiple contextual factors, including the user’s historical affinity for the target category and the temporal feasibility of such visits (e.g., bar visits in the morning are unlikely). Despite these constraints, this setting serves as a meaningful diagnostic for adaptability. Results in Table \ref{tab:preference-shift} indicate that GenUP successfully increases the predicted frequency of the target category across all four target categories, demonstrating its ability to promptly adapt its predictions in response to NL user profile edits. Crucially, this flexibility is achieved without any model fine-tuning, highlighting the efficiency of the NL user profile prompting framework. These findings underscore the potential of GenUP for real-time preference adaptation, enabling dynamic, user-controllable personalization in next POI recommendation systems.

\begin{figure}
    \centering
    \includegraphics[width=0.35\textwidth]{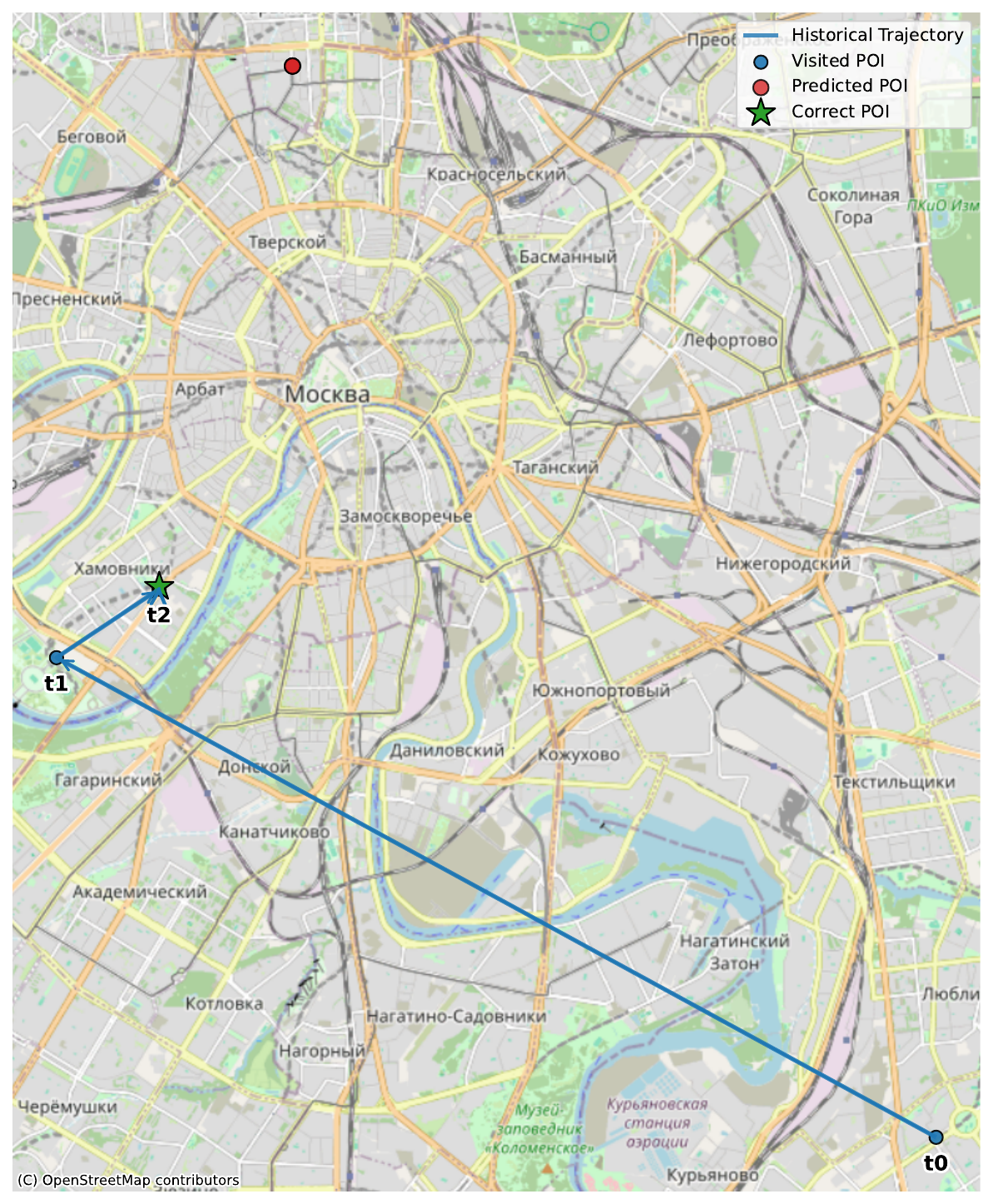}
    \caption{Example from the Moscow dataset where both GenUP and LLM4POI* incorrectly predicted a novel POI instead of the ground-truth revisit to a previously visited location within the same trajectory.}
    \Description{Example from the Moscow dataset where both GenUP and LLM4POI* incorrectly predicted a novel POI instead of the ground-truth revisit to a previously visited location within the same trajectory.}
    \label{fig:moscow-all-wrong}
\end{figure}

\begin{figure}
    \centering
    \includegraphics[width=0.35\textwidth]{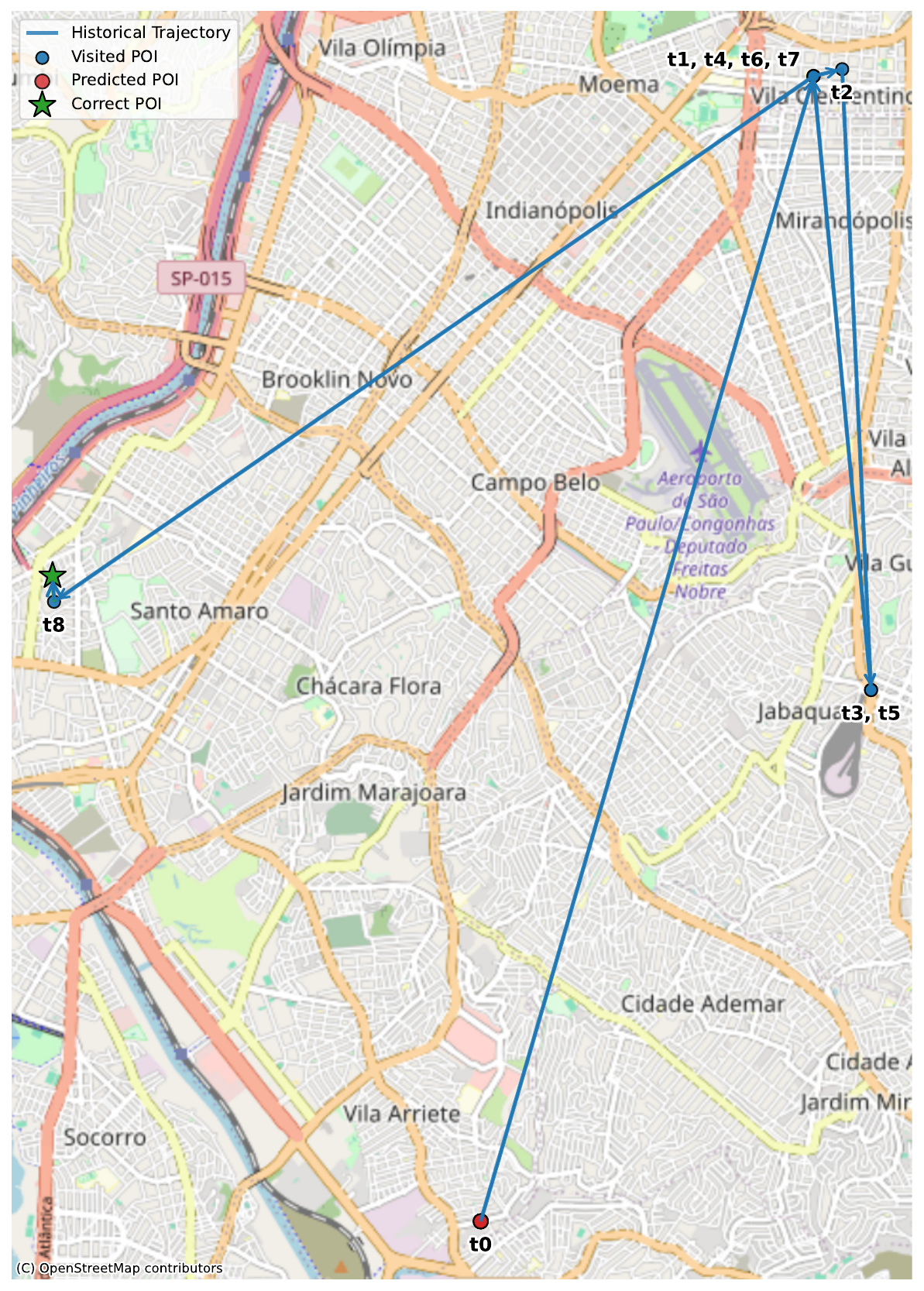}
    \caption{Example from the São Paulo dataset illustrating a failure case where both GenUP and LLM4POI* were overly conservative, predicting historically visited POIs instead of a plausible but novel ground-truth destination.}
    \Description{Example from the São Paulo dataset illustrating a failure case where both GenUP and LLM4POI* were overly conservative, predicting historically visited POIs instead of a plausible but novel ground-truth destination.}
    \label{fig:sao-paulo-all-wrong}
\end{figure}

\subsubsection{Qualitative Analysis}

To further investigate how GenUP compares to baseline models lacking user profiles, we conducted a qualitative analysis on the Moscow and São Paulo datasets. Specifically, we examined two representative cases: (1) instances where both GenUP and the LLM4POI* baseline incorrectly predicted the next POI, and (2) instances where all GenUP variants correctly predicted the next POI while LLM4POI* failed.

In the first scenario, we observed two types of systematic errors. In some cases, both GenUP and LLM4POI* were overly explorative, predicting a novel POI instead of returning to a previously visited location within the same trajectory. Figure \ref{fig:moscow-all-wrong} illustrates such a case from the Moscow dataset. Conversely, another failure mode emerged when the models were too conservative, overly reliant on historical preferences, and failed to explore plausible but novel POIs. This behavior is depicted in a São Paulo case shown in Figure \ref{fig:sao-paulo-all-wrong}. These observations highlight the inherent tension and tradeoff between exploiting routine patterns and exploring new but contextually relevant next POI options.

\begin{figure}
    \centering
    \hfill
    \begin{subfigure}[t]{0.215\textwidth}
        \centering
        \includegraphics[width=\textwidth]{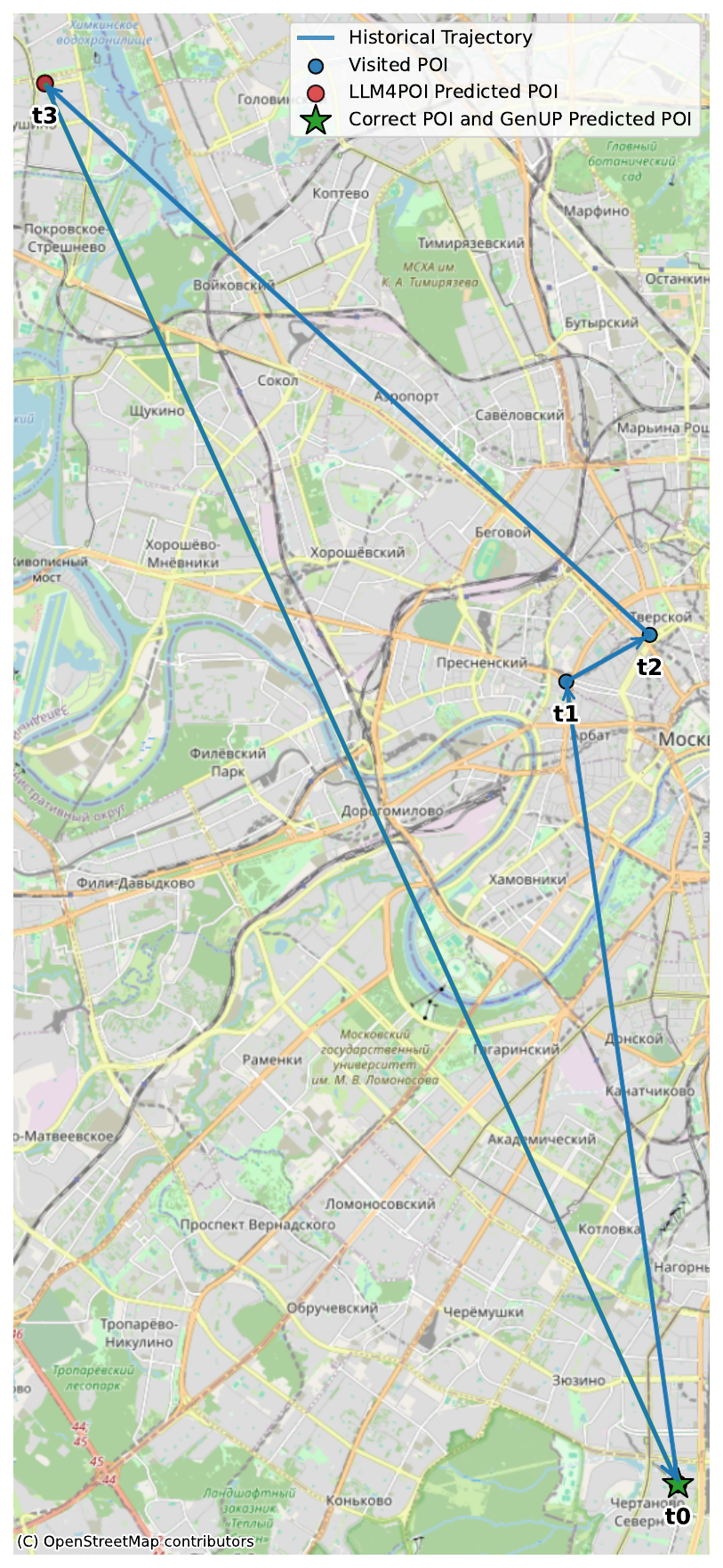}
        \caption{Trajectory in Moscow.}
    \end{subfigure}
    \hfill
    \begin{subfigure}[t]{0.21\textwidth}
        \centering
        \includegraphics[width=\textwidth]{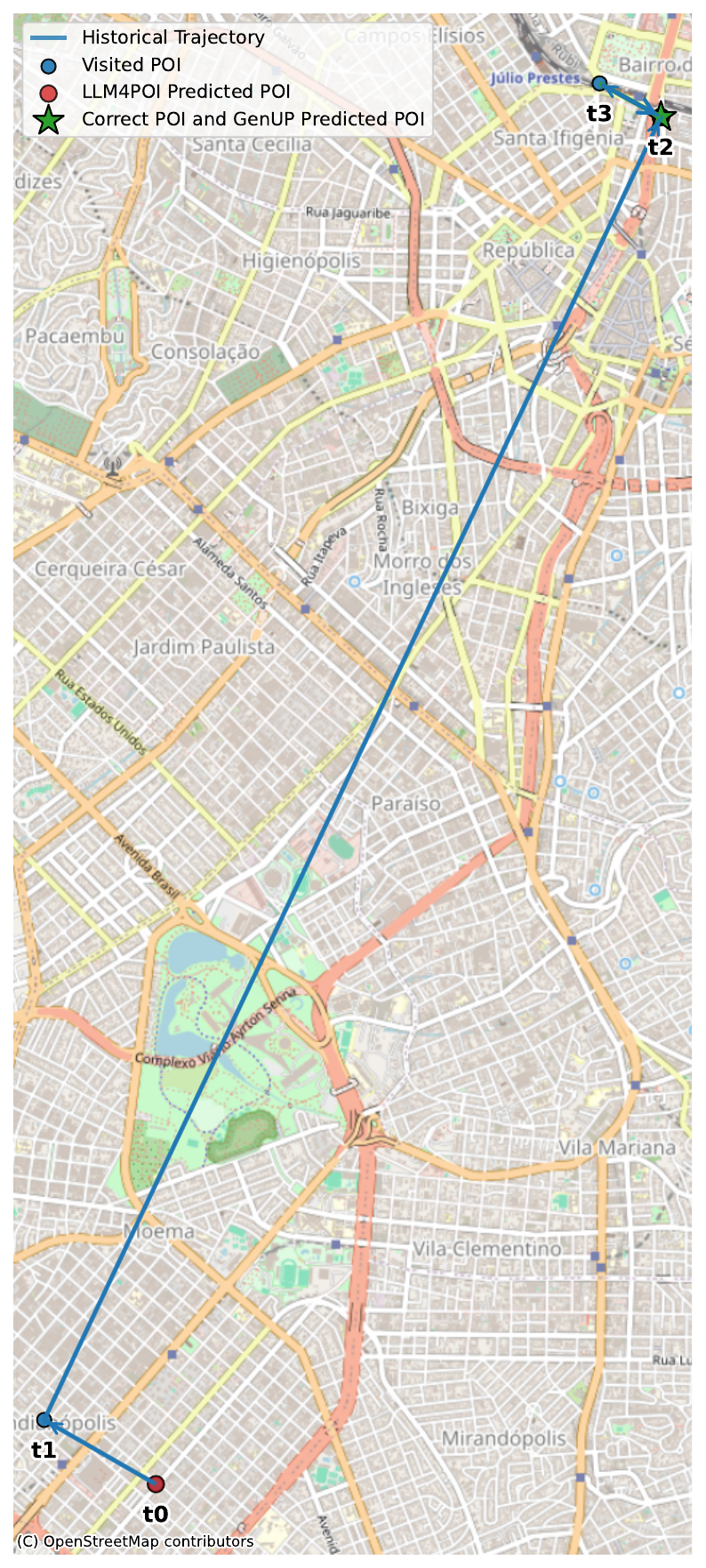}
        \caption{Trajectory in São Paulo.}
    \end{subfigure}
    \hfill
    \caption{Examples where all GenUP models correctly predicted the next POI, while LLM4POI did not. In both (a) Moscow and (b) São Paulo, the user returns to a previously visited POI, correctly captured by GenUP.}
    \Description{Examples where all GenUP models correctly predicted the next POI, while LLM4POI did not. In both (a) Moscow and (b) São Paulo, the user returns to a previously visited POI, correctly captured by GenUP.}
    \label{fig:genup-correct}
\end{figure}

In the second scenario, where all GenUP models correctly predicted the next POI and LLM4POI* failed, both models generated predictions that correspond to previously visited POIs within the current trajectory. However, only GenUP correctly identified the target POI that the user revisited.  As shown in Figure~\ref{fig:genup-correct}, which presents representative cases from the Moscow and São Paulo datasets, GenUP disambiguates among multiple historically plausible POIs by leveraging behavioral patterns contained in the natural language user profile. Whereas LLM4POI* operates solely on short-term trajectory context, GenUP conditions its predictions on a richer representation of long-term mobility behaviors. These examples highlight GenUP's advantage in routine-driven prediction scenarios, where personalized profiles inform the model with patterns and routines not found in the current trajectory alone.



\section{Conclusion and Future Work}

In this paper, we propose a novel framework for next POI recommendation by introducing generative natural language user profiles derived from LBSN check-in sequences. To the best of our knowledge, this is the first approach that generates NL user profiles directly from spatio-temporal, check-in data and integrates them into a next POI prediction framework. Our method addresses key challenges in POI recommendation, including cold-start scenarios and the lack of interpretability in embedding-based systems. By incorporating behavioral theories and user attributes, our profiles capture routines, preferences, and long-term mobility patterns to inform personalized POI recommendations. Our results demonstrate that our approach outperforms existing LLM-based baselines in terms of personalization, efficiency, and adaptability. Specifically, we observe consistent improvements in Acc@1 across multiple datasets, underscoring GenUP’s ability to better capture user-specific preferences. Our method achieves greater computational efficiency by eliminating the need for long input contexts and costly trajectory similarity computations required by prior work, such as LLM4POI. Moreover, we show that GenUP adapts to user preference shifts, where profiles can be updated with simple natural language edits without model re-training. Qualitative analyses further reveal that GenUP captures long-term user routines more effectively, enabling more accurate and context-aware predictions. Together, these findings highlight the promise of interpretable, editable NL user profiles to build more transparent and efficient LLM-based POI recommendation systems. In the future, we plan to introduce intent modeling via reasoning, which would provide greater transparency and further improve the accuracy of recommendations by more accurately capturing users' underlying motivations and decision-making processes.


\begin{acks}
We would like to thank the support of the National Computational Infrastructure and the ARC Center of Excellence for Automated Decision Making and Society (CE200100005). Computational facilities were provided by the School of Computer Science and Engineering at UNSW Sydney through the Wolfpack computational cluster. We further thank OpenAI’s Researcher Access Program for API access to GPT models and the Google Cloud Research Credits program for their support.
\end{acks}

\bibliographystyle{ACM-Reference-Format}
\bibliography{sample-sigconf}

\end{document}